\RequirePackage{fix-cm}
\documentclass[smallextended]{svjour3}       
\smartqed  
\usepackage{graphicx}
\input epsf
\expandafter\let\csname equation*\endcsname\relax
  \expandafter\let\csname endequation*\endcsname\relax
  \usepackage[fleqn]{amsmath}
\usepackage{amsfonts}
\usepackage{amssymb}
\usepackage{amsopn}
\usepackage{epsfig}
\usepackage{graphics,psfrag,rotating}
\usepackage{graphicx}
\usepackage{dcolumn}
\usepackage{bm}
\usepackage[numbers]{natbib}
\usepackage{epstopdf}
\usepackage{color}
\usepackage[usenames,dvipsnames,svgnames]{xcolor}
\usepackage[colorlinks=true,
            linkcolor=red,
            urlcolor=gray,
            citecolor=blue]{hyperref}
  \usepackage{hyperref}
\def \D {\tilde{\nabla}}

\def \ep {\varepsilon}

\def\ga{\gamma}

\def\div {\mbox{div}\,}

\def\sig{\sigma}

\def\om{\omega}

\def\3nab{\tilde{\nabla}}

\def\p{\partial}

\def\la {\langle}
\def\ra {\rangle}

\def\div{\mbox{div}}
\def\tl{\tilde}
\def\hsp5{\hspace{5mm}}
\newcommand{\sfrac}[2]{{\textstyle{#1\over#2}}}
\def\case#1/#2{\textstyle\frac{#1}{#2}}

\def\ber {\begin{eqnarray}}
\def\eer {\end{eqnarray}}
\def\bea {\begin{eqnarray}}
\def\eea {\end{eqnarray}}

\def\bc {\begin{center}}
\def\ec {\end{center}}
\def\case#1/#2{\frac{#1}{#2}}

\newcommand{\bw}{\begin{widetext}}
\newcommand{\ew}{\end{widetext}}
\newcommand{\nn}{\nonumber\\}

\newcommand{\lda}{\lambda}

\newcommand{\be}{\begin{equation}}
\newcommand{\bse}{\begin{subequation}}
\newcommand{\ese}{\end{subequation}}
\newcommand{\ee}{\end{equation}}
\newcommand{\eei}{\end{eqnarray}\indent\indent}
\newcommand{\bern}{\begin{eqnarray*}}
\newcommand{\eern}{\end{eqnarray*}}
\newcommand{\beast}{\begin{equation*}}
\newcommand{\eeast}{\end{equation*}}
\newcommand{\ba}{\begin{array}}
\newcommand{\ea}{\end{array}}
\newcommand{\bal}{\begin{eqnarray}}
\newcommand{\eal}{\end{eqnarray}}

\newcommand{\hs}{\,-\,}

\def\case#1/#2{\textstyle\frac{#1}{#2} }
\newcommand{\nb}{\nabla}

\newcommand{\bver}{\begin{verbatim}}
\newcommand{\ever}{\end{verbatim}}

\newcommand{\np}{\newpage}

\begin{document}

\title{Shear-free Anisotropic Cosmological Models in $f(R)$ Gravity
}


\author{Amare Abebe         \and
        Davood Momeni         \and
        Ratbay Myrzakulov 
        }


\institute{A. Abebe \at
             Department of Physics, North-West University,  Mahikeng 2735, South Africa \& Entoto Observatory and Research Center, P.O.Box 33679, Addis Ababa, Ethiopia \\
              \email{amare.abbebe@gmail.com}           
           \and
           D. Momeni \at
              Eurasian International Center for Theoretical Physics and Department of General \& Theoretical Physics, Eurasian National University, Astana 010008, Kazakhstan\\
               \email{davoodmomeni78@gmail.com}  
               \and
 R. Myrzakulov \at          
     Eurasian International Center for Theoretical Physics and Department of General \& Theoretical Physics, Eurasian National University, Astana 010008, Kazakhstan\\
         \email{rmyrzakulov@gmail.com}
}

\date{Received: date / Accepted: date}

\maketitle

\begin{abstract}
We study a class of  shear-free, homogeneous but anisotropic cosmological models with imperfect matter sources  in the context of $f(R)$ gravity.  We show that the anisotropic stresses are related to the electric part of the Weyl tensor in such a way that they balance each other. We also show that within the class of orthogonal $f(R)$ models, small perturbations of shear are damped, and that the electric part of the Weyl tensor and the anisotropic stress tensor decay with the expansion as well as the heat flux of the curvature fluid.
Specializing in locally rotationally symmetric spacetimes in orthonormal frames,  we examine the late-time behaviour of the de Sitter universe in $f(R)$ gravity. For the Starobinsky model of $f(R)$, we study the evolutionary behavior of the Universe  by numerically integrating the Friedmann equation, where the initial conditions for the expansion, acceleration and jerk parameters are taken from observational data.
\end{abstract}

\section{Introduction}
As a result of the current understanding that the Universe is in a state of accelerated expansion, many modifications to General Relativity (GR), the theory on which modern cosmology is based, have been proposed recently. One such modification consists of a class of higher-order gravity models that attempt to address the shortcomings of GR in the infrared (IR) and ultraviolet (UV) ranges \cite{capozziello11extended, modesto12, modesto14, biswas12, clifton12, de2010f, noj07,noj11}. These models  are generally obtained by including higher-order curvature invariants in the
Einstein-Hilbert action,  by making the action nonlinear in the Ricci curvature $R$, or contain terms involving combinations of derivatives of $R$, in which case the models are known as {\it  $f(R)$ theories of gravity}.

First proposed by Buchdal \cite{buchdal70}, $f(R)$ theories gained more popularity after further developments by Starobinsky \cite{staro80}  and later  following the realization of the  discrepancy between theory and observation \cite{carroll04,noj03,sotiriou07, clifton12, nojiri06,staro07}.

The role of shear in  general relativistic \cite{mimo93, ellis67, godel52,goldberg62,robinson63,ellis69, macc71, collins86, barrow77, macc70, koi11} and $f(R)$ cosmologies \cite{barr83,barr06,clift06,midd10,abebe11, abebe2013, amare14a,maye14} has been the subject of intense study for some time now,  with the studies focusing mostly on the special nature of shear-free cases. In particular, it was shown in \cite{mimo93} that in  the orthogonally spatially homogeneous models with vanishing  shear, the anisotropic stresses are related to the anisotropic curvature of the spatial hypersurface through the electric part of the Weyl tensor. It was also shown that within the class of orthogonal  models, small perturbations of shear are damped, and that the electric part of the Weyl tensor and the anisotropic stress tensor decay with the expansion.

The main focus of this work is the analysis of anisotropic but homogeneous, shear-free models whose underlying theory of gravitational interaction is $f(R)$-gravity.

The rest of this paper is organised as follows:  in Sec. \ref{covsec} a covariant description of $f(R)$  field equations is presented. In Sec. \ref{orth} we specialise to orthogonal cosmological models with anisotropic matter sources and analyse the properties of such models in the case of shear-free imperfect fluids in Sec. \ref{shearf}. In Sec. \ref{aniso}, the analysis is taken further by considering  subclasses of locally rotationally symmetric spacetimes with barotropic equations of state and a qualitative analysis of such models has been made.
Finally in Sec. \ref{conc} we discuss the results and give conclusions.


Natural units ($\hbar=c=k_{B}=8\pi G=1$)
will be used throughout this paper, and Latin indices run from 0 to 3.
The symbols $\nabla$, $\D$ and the overdot $^{.}$ represent the usual covariant derivative, the spatial covariant derivative, and differentiation with respect to cosmic time. We use the
$(-+++)$ spacetime signature and the Riemann tensor is defined by
\beast
R^{a}_{bcd}=\Gamma^a_{bd,c}-\Gamma^a_{bc,d}+ \Gamma^e_{bd}\Gamma^a_{ce}-
\Gamma^f_{bc}\Gamma^a_{df}\;,
\eeast
where the $\Gamma^a_{bd}$ are the Christoffel symbols (i.e., symmetric in
the lower indices), defined by
\beast
\Gamma^a_{bd}=\frac{1}{2}g^{ae}
\left(g_{be,d}+g_{ed,b}-g_{bd,e}\right)\;.
\eeast
The Ricci tensor is obtained by contracting the {\em first} and the
{\em third} indices of the Riemann tensor:
\beast
R_{ab}=g^{cd}R_{cadb}\;.
\eeast

The completely anti-symmetric pseudotensor $\eta^{abcd}$ is defined such that
\beast
\eta_{0123}=\sqrt{-g}\;,
\eeast
where $g=\det(g_{ab})$ is the determinant of the metric$g_{ab}$.

Unless otherwise stated, primes $^{'}$ etc are shorthands for derivatives with respect to the Ricci scalar
\beast
R=R^{a}{}_{a}\;
\eeast
and $f$ is used as a shorthand for $f(R)$.
Moreover the following standard notations are used:
\bern
&& (ab):~\mbox{symmetrization over the indices $a$ and $b$},\\
&& [ab]:~\mbox{anti-symmetrization over the indices $a$ and $b$},\\
&& \la ab\ra:~\mbox{orthogonal, symmetric, trace-free projection over the indices $a$ and $b$}.
\eern

\section{Covariant Description  of the Field Equations}\label{covsec}

In the  standard  $f(R)$-gravity formulation, one starts with the modified Einstein-Hilbert action 
\be
{\cal A}= \sfrac12 \int d^4x\sqrt{-g}\left[f(R)+2{\cal L}_m\right]\;,
\label{action}
\ee
 where  ${\cal L}_m$ stands for the matter field contribution to the Lagrangian,  and uses the variational principle of least action with respect to the metric $g_{ab}$ to obtain
the  generalised Einstein Field Equations (EFEs)
 \be
 G_{ab}=\tl T^{m}_{ab}+T^{R}_{ab}\equiv T_{ab}\;.
 \ee
 Here we have defined
 \be\label{emt}
\tl T^{m}_{ab}\equiv \frac{T^{m}_{ab}}{f'}\;,~~~~
T^{R}_{ab}\equiv \frac{1}{f'}\left[\sfrac{1}{2}(f-Rf')g_{ab}+\nb_{b}\nb_{a}f'-g_{ab}\nb_{c}
\nb^{c}f' \right]
\ee
as the effective matter and curvature energy-momentum tensors (EMTs), respectively.
The EMT of standard matter  is given by 
\be T^{m}_{ab} =\frac{2}{\sqrt{-g}}\frac{\delta(\sqrt{-g}{\cal{L}}_m)}{\delta g^{ab}}= \mu_{m}u_{a}u_{b} + p_{m}h_{ab}+ q^{m}_{a}u_{b}+ q^{m}_{b}u_{a}+\pi^{m}_{ab}\;,\ee
where $\mu_{m}$, $p_{m}$, $q^{m}_{a}$ and $\pi^m_{ab}$ are the associated energy density, 
isotropic pressure, heat flux and anisotropic pressure, respectively, and  $u^a\equiv \frac{dx^a}{dt}$  is the normalized  $4$-velocity of fundamental observers comoving with the fluid. We use this  vector to define the  covariant time derivative for any tensor 
${S}^{a..b}_{c..d} $ along an observer's worldlines:
\be
\dot{S}^{a..b}_{c..d}{} = u^{e} \nb_{e} {S}^{a..b}_{c..d}\;.
\ee
On the other hand, we use  the projection tensor $ h_{ab}\equiv g_{ab}+u_au_b$ to define the fully orthogonally projected covariant derivative  for any tensor ${S}^{a..b}_{c..d} $:
\be
\tl\nb_{e}S^{a..b}_{c..d}{} = h^a_f h^p_c...h^b_g h^q_d 
h^r_e \nb_{r} {S}^{f..g}_{p..q}\;,
\ee
with total projection on all the free indices. We extract the orthogonally projected symmetric trace-free part of vectors and  rank-2 tensors using
\be
V^{\langle a \rangle} = h^{a}_{b}V^{b}~, ~ S^{\langle ab \rangle} = \left[ h^{(a}_c {} h^{b)}_d 
- \sfrac{1}{3} h^{ab}h_{cd}\right] S^{cd}\;,
\label{PSTF}
\ee
and the  volume element for the restspaces orthogonal to $u^a$ is given by \cite{Ellis98}
\be
\ep_{abc}=u^{d}\eta_{dabc}=-\sqrt{|g|}\delta^0_{\left[ a \right. }\delta^1_b\delta^2_c\delta^3_{\left. d \right] }u^d\Rightarrow \ep_{abc}=\ep_{[abc]},~\ep_{abc}u^{c}=0,
\ee
where $\eta_{abcd}$ is the 4-dimensional volume element satisfying the conditions
\be
\eta_{abcd}=\eta_{[abcd]}=2\ep_{ab[c}u_{d]}-2u_{[a}\ep_{b]cd}.
\ee
The covariant spatial divergence and curl of vectors and rank-2 tensors  are given as \cite{maartens97}
\ber
&& \div V=\D^aV_a\,,~~~~~~(\div S)_a=\D^bS_{ab}\,, \\
&& curl V_a=\ep_{abc}\D^bV^c\,,~~ curl S_{ab}=\ep_{cd(a}\D^cS_{b)}{}^d \,.
\eer
 The 4-velocity vector field $u^a$ can be split  into its
irreducible parts as follows
\be\label{congruence}
\nb_au_b=-A_au_b+\sfrac13h_{ab}\Theta+\sigma_{ab}+\ep_{a b c}\omega^c,
\ee
where $A_a\equiv \dot{u}_a$, $\Theta\equiv \tl\nb_au^a$, 
$\sigma_{ab}\equiv \tl\nb_{\langle a}u_{b \rangle}$ and $\omega^{a}\equiv\ep^{a b c}\tl\nb_bu_c\;.$\\
We can also split the  Weyl  conformal curvature tensor \cite{Ellis98, betschart}
\be
C^{ab}{}_{cd}=R^{ab}{}_{cd}-2g^{[a}{}_{[c}R^{b]}{}_{d]}+\frac{R}{3}g^{[a}{}_{[c}g^{b]}{}_{d]}
\ee
into its ``gravito-electric'' (GE) and ``gravito-magnetic'' (GM) parts, respectively, as
\be
E_{ab}\equiv C_{agbh}u^{g}u^{h},~~~~~~~H_{ab}=\sfrac{1}{2}\eta_{ae}{}^{gh}C_{ghbd}u^{e}u^{d}.
\ee
The GE and GM components represent the free gravitational field \cite{Ellis98} and they describe gravitational action at a distance \hs tidal forces and gravitational waves. They influence the motion of matter and radiation through the geodesic deviation for timelike and null-vector fields, respectively. 

The total energy density, isotropic and anisotropic pressures  and heat flux of the $f(R)$ universe are given, respectively,  by \cite{carloni08}
\be\label{totaltherm}
\mu\equiv\frac{\mu_{m}}{f'}+\mu_{R}\;,~~~\;p\equiv\frac{p_{m}}{f'}+p_{R}\;,~~~~~~\;\pi_{ab}\equiv\frac{\pi^{m}_{ab}}{f'}+\pi^{R}_{ab}\;,~~~
q_{a}\equiv \frac{q^{m}_{a}}{f'}+q^{R}_{a}\;,
\ee
where the  thermodynamic quantities for the {\it curvature fluid} component  are defined as
\ber
&&\label{mur}\mu_{R}=\frac{1}{f'}\left[\frac{1}{2}(Rf'-f)-\Theta f'' \dot{R}+ f''\tilde{\nabla}^{2}R  \right]\;,\\
&&\label{pr}p_{R}=\frac{1}{f'}\left[\frac{1}{2}(f-Rf')+f''\ddot{R}+f'''\dot{R}^{2}\right.\nn
&&\left.~~~~~~~~~~~~+\frac{2}{3}\left( \Theta f''\dot{R}-f''\tilde{\nabla}^{2}R -f'''\tilde{\nabla}^{a}R \tilde{\nabla}_{a}R\right)  \right]\;,\\
&&\label{qar}q^{R}_{a}=-\frac{1}{f'}\left[f'''\dot{R}\tilde{\nabla}_{a}R +f''\tilde{\nabla}_{a}\dot{R}-\frac{1}{3}f''\Theta \tilde{\nabla}_{a}R \right]\;, \\ 
&&\label{pir} \pi^{R}_{ab}=\frac{1}{f'}\left[f''\tilde{\nabla}_{\langle a}\tilde{\nabla}_{b\rangle}R +f'''\tilde{\nabla}_{\langle a}R\tilde{\nabla}_{b\rangle}R-\sigma_{ab}\dot{R} f''\right]\;. 
\eer
 
In the $1+3$ covariant decomposition \cite{ellis89,ellis12}, a fundamental observer slices spacetime into time and space.  The Bianchi and Ricci identities
\be\label{biricci}
\nb_{[a}R_{bc]d}{}^{e}=0\;,~~~~
(\nb_{a}\nb_{b}-\nb_{b}\nb_{a})u_{c}=R_{abc}{}^{d}u_{d}\;
\ee
applied on the total fluid 4-velocity $u^{a}$ result in evolution equations\hs which propagate consistent initial data on some  initial ($t=t_{0}$) hypersurface $S_{0}$  uniquely along  timelike congruences\hs and constraint equations\hs which restrict the initial data to be specified \cite{maartens98}. In $f(R)$ gravity, the evolution equations are given by \cite{carloni08}
\ber
&&\label{mue}\dot{\mu}_{m}=-(\mu_{m}+p_{m})\Theta-\tl\nb^{a}q^{m}_{a}-2A_{a}q^{a}_m-\sigma^{a}_{b}\pi^{b}_{a,m}\;,\\
&&\dot{\mu}_{R}=-(\mu_{R}+p_{R})\Theta+\frac{\mu_{m}f''}{f'^{2}}\dot{R}-\D^{a}q^{R}_{a}-2A_{a}q^{a}_R-\sigma^{a}_{b}\pi^{b}_{a,R}\;,\\
&&\label{ray}\dot{\Theta}=-\sfrac13 \Theta^2-\sfrac12(\mu+3p)+\tl\nb_aA^a-A_aA^a-\sigma_{ab}\sigma^{ab}+2\omega_{a}\omega^{a}\;,\\
&&\label{qe}\dot{q}^{m}_{a}=-\sfrac{4}{3}\Theta q^{m}_{a}-(\mu_{m}+p_m)A_{a}-\D_a p_m-\D^{b}\pi^{m}_{ab}-\sigma^{b}_{a}q^{m}_b-A^{b}\pi^m_{ab}-\ep_{abc}\omega^{b}q^{c}_m\;,\\
&&\label{qer}\dot{q}^{R}_{a}=-\sfrac{4}{3}\Theta q^{R}_{a}+\frac{\mu_{m}f''}{f'^{2}}\D_{a}R-\D_{a}p_{R}-\D^{b}\pi^{R}_{ab}-\sigma^{b}_{a}q^{R}_b\nn
&&~~~~~~~-(\mu_R+p_R)A_{a}-A^{b}\pi^R_{ab}-\ep_{abc}\omega^{b}q^{c}_R\;,\\
&&\label{propom}\dot{\omega}_{a}=-\sfrac23\Theta\omega_{a}-\sfrac{1}{2}\ep_{abc}\tl\nb^{b}A^{c}+\sigma^{b}_{a}\omega_{b}\;,\\
&&\label{sig}\dot{\sigma}_{ab}=-\sfrac{2}{3}\Theta\sigma_{ab}-E_{ab}+\sfrac{1}{2}\pi_{ab}+\tl\nb_{\la a}A_{b\ra}+A_{\langle a}A_{b\rangle}-\sigma^c_{\langle }{}_{a}\sigma_{b\rangle c}-\omega_{\langle a}\omega_{b\rangle}\;,\\
&&\label{gep}\dot{E}_{ab}+\sfrac{1}{2}\dot{\pi}_{ab}=\ep_{cd\langle a}\tl\nb^{c}H_{b\rangle }^{d}-\Theta \left(E_{ab}+\sfrac{1}{6}\pi_{ab}\right)-\sfrac{1}{2}\left(\mu+p\right)\sigma_{ab}
-\sfrac{1}{2}\tl\nb_{\langle a}q_{b\rangle}\nn
&&~~~~~+3\sigma^{\langle c}_{a}\left(E_{b\rangle c}-\sfrac{1}{6}\pi_{b\rangle c}\right)-A_{\langle a}q_{b\rangle}+\ep_{cd\langle a}\left[2A^{c}H^d_{b\rangle}+\omega^{c}(E^d_{b\rangle}+\sfrac{1}{2}\pi^d_{b\rangle})\right]\;,\\
&&\label{gmp}\dot{H}_{ab} =-\Theta H_{ab}-\ep_{cd\langle a}\tl\nb^{c}E_{b\rangle }^{d}+
\sfrac{1}{2}\ep_{cd\langle a}\tl\nb^{c}\pi^{d}_{b\rangle}\nn
&&~~~~~~~~+3\sigma^{\langle c}_{a}H_{b\rangle c}+\sfrac{3}{2}\omega_{\langle a}q_{b\rangle}-\ep_{cd\langle a}\left[2A^{c}E^d_{b\rangle}-\sfrac{1}{2}\sigma^c_{b\rangle}q^{d}-\omega^{c}H^{d}_{b\rangle}\right]\;,
\eer
whereas the constraints read
\ber
&&\label{R4} (C^{1})_{a}:=\D^{b}\sigma_{ab}-\sfrac{2}{3}\tl\nb_{a}\Theta+\ep_{abc}\left(\tl\nb^{b}\omega^{c}+2A^{b}\omega^{c}\right)+q_{a}=0\;,\\
&&\label{R6} (C^{2})_{ a b}:=\ep_{cd(a}\tl\nb^{c}\sigma_{b)}{}^{d}+\tl\nb_{\langle a}\omega_{b \rangle}-H_{a b}-2A_{\langle a}\omega_{b\rangle}=0\;,\\
&&\label{B6} (C^{3})_{a}:=\tl\nb^{b}H_{ab}+(\mu+p)\omega_{a}+\ep_{abc}\left[\sfrac12\tl\nb^{b}q^{c}+\sigma_{bd}\left(E^{d}{}_{c}+\sfrac{1}{2}\pi^{d}{}_{c}\right)\right]\nn
&&~~~~~~~~~~~+3\omega_{b}\left(E^{ab}-\sfrac{1}{6}\pi^{ab}\right)=0\;,\\
&&\label{B5} (C^{4})_{a}:=\tl\nb^{b}E_{ab}+\sfrac{1}{2}\tl\nb^{b}\pi_{ab}-\sfrac13\tl\nb_{a}\mu+
\sfrac13\Theta q_{a}\nn
&&~~~~~~~~~~~~-\sfrac{1}{2}\sigma^{b}_{a}q_{b}-3\omega^b H_{ab}-\ep_{abc}[\sigma^{bd}H^{c}_{d}-\sfrac{3}{2}\omega^{b}q^{c}]=0\;,\\
&& \label{R5} (C^{5}):=\tl\nb^a\omega_a-A_a\om^a=0\;.
\eer
The Gau{\ss}-Codazzi equations are given by
\be\label{gauss}
\tl R_{ab}+\dot{\sig}_{\la ab\ra}+\Theta\sig_{ab}-\D_{\la a}A_{b\ra}-A_{\la a}A_{b\ra}-\pi_{ab}-\frac{1}{3}\left(2\mu-\frac{2}{3}\Theta^{2}\right)h_{ab}=0\;,
\ee
where $\tl R_{ab}$ is the Ricci tensor on 3-D spatial hypersurfaces with $\tl R=2\mu-\frac{2}{3} \Theta^{2}+2\sig^{2}$ as its corresponding (3-curvature) Ricci scalar.
\section{Orthogonal Models}\label{orth}
Following \cite{mimo93}, the orthogonal models are characterised by the matter EMT representing an anisotropic fluid without heat fluxes
\be T^{m}_{ab} = \mu_{m}u_{a}u_{b} + p_{m}h_{ab}+\pi^{m}_{ab}\;,\ee
 the matter energy density and isotropic pressure
measured by an observer moving with the velocity $u^a$.
 In this setting, we have  an irrotational and non-accelerated flow of the vector field $u^a$ and therefore $\om_a=0=A_a$. Thus the corresponding evolution and constraint equations are given by
 \ber
&&\label{mueo}\dot{\mu}_{m}=-(\mu_{m}+p_{m})\Theta-\sigma^{a}_{b}\pi^{b}_{a,m}\;,\\
&&\label{muro}\dot{\mu}_{R}=-(\mu_{R}+p_{R})\Theta+\frac{\mu_{m}f''}{f'^{2}}\dot{R}-\D^{a}q^{R}_{a}-\sigma^{a}_{b}\pi^{b}_{a,R}\;,\\
&&\label{rayo}\dot{\Theta}=-\sfrac13 \Theta^2-\sfrac12(\mu+3p)-\sigma_{ab}\sigma^{ab}\;,\\
&&\label{qero}\dot{q}^{R}_{a}=-\sfrac{4}{3}\Theta q^{R}_{a}+\frac{\mu_{m}f''}{f'^{2}}\D_{a}R-\D_{a}p_{R}-\D^{b}\pi^{R}_{ab}-\sigma^{b}_{a}q^{R}_b\;,\\
&&\label{sigo}\dot{\sigma}_{ab}=-\sfrac{2}{3}\Theta\sigma_{ab}-E_{ab}+\sfrac{1}{2}\pi_{ab}-\sigma^c_{\langle }{}_{a}\sigma_{b\rangle c}\;,\\
&&\label{gepo}\dot{E}_{ab}+\sfrac{1}{2}\dot{\pi}_{ab}=\ep_{cd\langle a}\tl\nb^{c}H_{b\rangle }^{d}-\Theta \left(E_{ab}+\sfrac{1}{6}\pi_{ab}\right)-\sfrac{1}{2}\left(\mu+p\right)\sigma_{ab}
-\sfrac{1}{2}\tl\nb_{\langle a}q^R_{b\rangle}\nn
&&~~~~~~~~~~~~~~~~~+3\sigma^{\langle c}_{a}\left(E_{b\rangle c}-\sfrac{1}{6}\pi_{b\rangle c}\right)\;,\\
&&\label{gmpo}\dot{H}_{ab} =-\Theta H_{ab}-\ep_{cd\langle a}\tl\nb^{c}E_{b\rangle }^{d}+
\sfrac{1}{2}\ep_{cd\langle a}\tl\nb^{c}\pi^{d}_{b\rangle}+3\sigma^{\langle c}_{a}H_{b\rangle c}\nn
&&~~~~~~~~~+\sfrac{1}{2}\ep_{cd\langle a}\sigma^c_{b\rangle}q^{d}_R\;,
\eer

\ber
&&\label{R4o} (C^{\ast 1})_{a}:=\D^{b}\sigma_{ab}-\sfrac{2}{3}\tl\nb_{a}\Theta+q^R_{a}=0\;,\\
&&\label{R6o} (C^{\ast 2})_{ a b}:=\ep_{cd(a}\tl\nb^{c}\sigma_{b)}{}^{d}-H_{a b}=0\;,\\
&&\label{B6o} (C^{\ast 3})_{a}:=\tl\nb^{b}H_{ab}+\ep_{abc}\left[\sfrac12\tl\nb^{b}q^{c}_R+\sigma_{bd}\left(E^{d}{}_{c}+\sfrac{1}{2}\pi^{d}{}_{c}\right)\right]=0\;,\\
&&\label{B5o} (C^{\ast 4})_{a}:=\tl\nb^{b}E_{ab}+\sfrac{1}{2}\tl\nb^{b}\pi_{ab}-\sfrac13\tl\nb_{a}\mu+
\sfrac13\Theta q^R_{a}-\sfrac{1}{2}\sigma^{b}_{a}q_{b}\nn
&&~~~~~~~~~~~~~-\ep_{abc}\sigma^{bd}H^{c}_{d}=0\;,
\eer
where the new equations corresponding to Eqns \eqref{propom} and \eqref{R5} become trivial, and with Eqn \eqref{qe} resulting in the constraint 
\be\label{ppic} (C^{\ast 5})_{a}:=\D_a p_m+\D^{b}\pi^{m}_{ab}=0\;.\ee 
\section{Shear-free Anisotropic Models with an Imperfect Fluid}\label{shearf}
From causal relativistic thermodynamical relationships for imperfect fluids, the anisotropic pressure is known to evolve according to \cite{wern76,novella96,maartens96,rezz13}
\be
\tau \dot{\pi}_{ab}+\pi_{ab}=-\lambda\sigma_{ab}\;,
\ee
where $\tau$ and $\lda$ are relaxation and viscosity parameters. If we consider cases where $\tau$ is negligible and $\lda$ is a positive constant, and use the fairly popular ansatz (valid near thermal equilibrium, such as in the very early stages of the Universe) for the equation of state \cite{macc70,mimo93, col95}
\be\label{eoshear}
\pi_{ab}=-\lda \sig_{ab}\;,
\ee
 then Eqns \eqref{totaltherm} and \eqref{pir} imply that we can rewrite \eqref{eoshear} as
\be\label{preshear}
\pi^{m}_{ab}+f''\tilde{\nabla}_{\langle a}\tilde{\nabla}_{b\rangle}R +f'''\tilde{\nabla}_{\langle a}R\tilde{\nabla}_{b\rangle}R=\sigma_{ab}\left(\dot{R} f''-\lda f'\right)\;.
\ee
For a general case of vanishing shear tensor during the entire cosmic evolution, one can see from Eqn \eqref{preshear} that
\be
\pi^{m}_{ab}=-f''\tilde{\nabla}_{\langle a}\tilde{\nabla}_{b\rangle}R -f'''\tilde{\nabla}_{\langle a}R\tilde{\nabla}_{b\rangle}R\;.
\ee
Moreover, the Gau{\ss}-Codazzi equations \eqref{gauss} reduce to
\be\label{gausss}
\tl R_{ab}-\sfrac{1}{3}\tl Rh_{ab}=\pi_{ab}=\frac{1}{f'}\left(\pi^{m}_{ab}+f''\tilde{\nabla}_{\langle a}\tilde{\nabla}_{b\rangle}R +f'''\tilde{\nabla}_{\langle a}R\tilde{\nabla}_{b\rangle}R\right)\;,
\ee
thus showing that even if the matter anisotropic stress vanishes, no constant-curvature geometries are guaranteed and hence no necessarily FLRW universes. It is also worth noticing that, unlike in GR, if we allow the matter anisotropic pressure to be nonzero despite a vanishing shear, constant-curvature models are allowed provided
\be\label{constcurv}
f''\tilde{\nabla}_{\langle a}\tilde{\nabla}_{b\rangle}R +f'''\tilde{\nabla}_{\langle a}R\tilde{\nabla}_{b\rangle}R=0\;.
\ee
The converse also holds, i.e., it is possible, unlike in GR, to have a vanishing matter anisotropic pressure $\pi^{m}_{ab}$ for a non-constant  curvature geometry.

One can see the tidal effect  on the anisotropic stresses by dropping the shear terms of Eqn \eqref{sigo}, obtaining the equation
\be\label{piel}
\pi_{ab}=2E_{ab}\;,
\ee
which shows that, in this case as in GR \cite{mimo93}, the anisotropic stresses are related to the electric
part of the Weyl tensor in such a way that they balance each other,  a necessary and sufficient condition
for the shear to remain zero if initially vanishing.

If the shear is nonzero, but with very small second-order contributions, then one can show that Eqn \eqref{sigo} can be approximated by
\be
\label{sigo2}\dot{\sigma}_{ab}\approx -\sfrac{2}{3}\Theta\sigma_{ab}\;.
\ee
Rewriting Eqn \eqref{sigo2} as
\be
\left(\sig^{2}\right)^{.}\approx -\sfrac{4}{3}\Theta\sig^{2}
\ee
shows that the shear decays with expansion. One can, therefore, conclude that
 \emph{within the class of orthogonal f(R) models, small perturbations of shear
are damped}, i.e. that these models are stable if expanding, a result similar to that obtained in \cite{mimo93} for models whose underlying theory is GR.

For shear-free orthogonal models satisfying Eqn \eqref{piel}, we see that Eqn \eqref{R6o} implies a purely electric Weyl tensor, i.e., $H_{ab}=0$, and hence Eqn \eqref{gmpo} reduces to an identity:
\be
\ep_{cd\langle a}\tl\nb^{c}E_{b\rangle }^{d}=
\sfrac{1}{2}\ep_{cd\langle a}\tl\nb^{c}\pi^{d}_{b\rangle}\;.\ee
Moreover, it is straighforward to show using Eqns \eqref{gepo} and \eqref{B5o} that
\ber
&&\label{gepo2}\dot{E}_{ab}=-\sfrac{2}{3}\Theta E_{ab}
-\sfrac{1}{4}\tl\nb_{\langle a}q^R_{b\rangle}\;,\\
&&\label{B5o2} \tl\nb^{b}E_{ab}=\sfrac16\left(\tl\nb_{a}\mu-\sfrac13\Theta q^R_{a}\right)\;.
\eer
Defining $E^{2}\equiv E_{ab}E^{ab}$, we can rewrite Eqn \eqref{gepo2} as
\be
\left(E^{2}\right)^{.}=-\sfrac{4}{3}\Theta E^{2}-\frac{1}{8}\left(\tl\nb_{\langle a}q^R_{b\rangle}E^{ab}+\tl\nb^{\langle a}q^{b\rangle}_{R}E_{ab}\right)\;,
\ee
thus showing the decay of the electric part of the Weyl tensor and
the anisotropic stress tensor  with the expansion. This equation  also implies decay  with the heat flux of the curvature fluid if the bracketed terms in the r.h.s are overall positive.

Let us now consider the generalized Friedman equation
\be
\Theta^{2}=3\left(\mu-\sfrac{1}{2}\tl R\right)\;.
\ee 
Since the total energy density $\mu$ is not  always guaranteed to be positive for generic $f(R)$ models, it is not straightforward to comment on the asymptotic isotropization of expanding shear-free anisotropic models for the different values of the spatial curvature. This is in contrast to the GR result where, for example, expanding shear-free  models which exhibit negative spatial curvature asymptotically approach isotropy \cite{mimo93}.
\section{Anisotropic LRS Models}\label{aniso}
Let us consider the locally rotationally symmetric (LRS) metric given by
\be
ds^{2}=-dt^{2}+a^{2}(t)dr^{2}+b^{2}(t)\left[d\theta^{2}+f^{2}(\theta)d\phi^{2}\right]\;,
\ee
where 
\[
f(\theta)=
\left\{
\begin{array}{ll}
 \sin(\theta)&\mbox{for $\tl R>0$}~ \mbox{(Kantowski-Sachs)},\\
\theta&\mbox{for $\tl R=0$}~ \mbox{(Bianchi I)},\\
 \sinh(\theta)&\mbox{for $\tl R<0$}~\mbox{(Bianchi III)}.
\end{array}
\right.
\]
Here $\tl R=2k/b^{2}$ for $k=\pm 1, 0$.
The non-vanishing kinematic quantities for these models are the expansion and shear, respectively given as
\ber
&&\Theta=\frac{\dot{a}}{a}+2\frac{\dot{b}}{b}\;,\\
&&\label{shearab} 2\sigma^{2}=\sigma_{ab}\sigma^{ab}=\frac{1}{\sqrt{3}}\left(\frac{\dot{a}}{a}-\frac{\dot{b}}{b}\right)\;.
\eer
Consider the EMT of the imperfect fluid matter source  to be of the form
\be
T_{ab}=\mu u_{a}u_{b}+\bar{p}h_{ab}-\bar{\pi}(e_{1})_{a}(e_{1})_{b}
\ee
where, because of the rotational symmetry, $e_{1}=\frac{1}{a}\frac{\p}{\p r}$ is defined as the unit vector along the axis of symmetry. Whereas $\mu$ represents the total energy density measured by a comoving observer, the pressure measured by the same observer is  
\be
p=\bar{p}-\bar{\pi}\;.
\ee
Here the anisotropic stress tensor in the orthonormal tetrad bases
\be
e_{0}=\frac{\p}{\p t}\;,~e_{1}=\frac{1}{a}\frac{\p}{\p r}\;,~e_{2}=\frac{1}{b}\frac{\p}{\p \theta}\;,~e_{3}=\frac{1}{b\sin\theta}\frac{\p}{\p \phi}
\ee
is given by
\be
\pi_{ab}=diag\left(0,-\frac{2}{3}\bar{\pi},\frac{1}{3}\bar{\pi},\frac{1}{3}\bar{\pi}\right)\;.
\ee
This way we can write the modified  EFEs as
\ber
&&\label{efe1}2\frac{\dot{a}\dot{b}}{ab}+\frac{k+\dot{b}^{2}}{b^{2}}=\mu\;,\\
&&\label{efe2}2\frac{\ddot{b}}{b}+\frac{k+\dot{b}^{2}}{b^{2}}=-\bar{p}+\bar{\pi}\;,\\
&&\label{efe3}\frac{\ddot{a}}{a}+\frac{\ddot{b}}{b}+\frac{\dot{a}\dot{b}}{ab}=-\bar{p}\;,
\eer
whereas the conservation equations \eqref{mueo}, \eqref{muro}, \eqref{qero} and \eqref{ppic} are rewritten as
\ber
&&\label{mueos}\dot{\mu}_{m}=-\left(\mu_{m}+\bar{p}_{m}-\sfrac13\bar{\pi}_m\right)\Theta\;,\\
&&\label{muros}\dot{\mu}_{R}=-\left(\mu_{R}+\bar{p}_{R}-\sfrac13\bar{\pi}_R\right)\Theta+\frac{\mu_{m}f''}{f'^{2}}\dot{R}-\D^{a}q^{R}_{a}\;,\\
&&\label{qeros}\dot{q}^{R}_{a}=-\sfrac{4}{3}\Theta q^{R}_{a}+\frac{\mu_{m}f''}{f'^{2}}\D_{a}R-\D_{a}\bar{p}_{R}-\D_a\bar{\pi}_{R}\;,\\
&&\label{dpmos}\D_a\bar{p}_m(e_1)^a=\D_a\bar{\pi}_m(e_1)^a\;.
\eer
As a result of the homogeneity assumption, $\bar{\pi}=\bar{\pi}_m(t)$ and therefore Eqn\eqref{dpmos} is trivially satisfied.

We notice from Eqn \eqref{shearab} that for the case of vanishing shear, $a(t)=b(t)$ and thus the modified EFES \eqref{efe1}-\eqref{efe3} reduce to
\ber
&&\label{efe11}3\frac{\dot{a}^2}{a^2}+\frac{k}{a^{2}}=\mu\;,\\
&&\label{efe22}2\frac{\ddot{a}}{a}+\frac{\dot{a}^{2}}{a^{2}}+\frac{k}{a^{2}}=-\bar{p}+\bar{\pi}\;,\\
&&\label{efe32}2\frac{\ddot{a}}{a}+\frac{\dot{a}^2}{a^2}=-\bar{p}\;.
\eer
Subtracting Eqn \eqref{efe32} from Eqn \eqref{efe22} yields
\be
\bar{\pi}=\frac{k}{a^{2}}\;,
\ee
and therefore
\be
E_{ab}=diag\left(0,-2E,E,E\right)\;,
\ee
where $E=\frac{\bar{\pi}}{6}$.

We adopt the barotropic EoS , $p_m=(\ga_m-1)\mu_m$, where $p_m=\bar{p}_m-\bar{\pi}_m/3$, from the continuity Eqn (\ref{mueos}) for $p_{m}$, we obtain $\mu_m=\mu_m^{0}a^{-3\ga_m}$. To integrate \eqref{efe11} we need to know $\mu_R$. Indeed, it is a hard job to integrate \eqref{muros} although we are working in the homogeneous case. But we can rewrite (\ref{efe11})  in the following form:
\ber
\label{efe111}3\frac{\dot{a}^2}{a^2}+\frac{k}{a^{2}}=\mu_m^{0}a^{-3\ga_m}+\frac{1}{f'}\left[\frac{1}{2}(Rf'-f)-\frac{3\dot{a}}{a} f'' \dot{R} \right]\;.
\eer
Here $\mu_m^0$ is the matter density at the time $t=t_0$ and $\ga_m$ is the EoS parameter for the matter content. As we see, Eqn \eqref{efe111} is model dependent. To specify solutions we must choose a specific model of $f(R)$ gravity. Otherwise, we cannot integrate it explicitly. Let us have a brief qualitative analysis of \eqref{efe111}. 
If we are looking for the late-time behavior of the solutions for \eqref{efe111} and if we suppose that the space is flat $k=0$, and without matter, the evolution is defined by the  de Sitter (dS) solution, in which  we put $R=6H_0^2$, where $H_0$ is the time scale of the dS universe. In this simple case, we can solve Eqn \eqref{efe111} to obtain:
\ber
\label{dS}H_0^2=\frac{1}{6f'}(Rf'-f).
\eer
But this is not the only case  we can solve \eqref{efe11}. Suppose that we choose a model of $f(R)$, so \eqref{efe11} reduces generally to a fourth-order ODE, which can be solved in terms of quadratures. For example, in the so-called Starobinsky model, $f(R)=R+\alpha R^2$, which is motivated for the inflationary universe  scenario  \cite{staro80}, Eqn \eqref{efe111} reduces to the following differential equation:
\ber
\label{efe1-1}3\frac{\dot{a}^2}{a^2}=\mu_m^{0}a^{-3\ga_m}+\frac{\alpha}{2}\frac{R^2-12H\dot{R}}{1+2\alpha R}\;,
\eer
where $R=6\left(\frac{\dot{a}^2}{a^2}+\frac{\ddot{a}}{a}\right)$. Eqn (\ref{efe1-1} ) is a third oder ODE for $a(t)$. So we need to specify initial condition(s) (ICs), as well as integrability condition(s). The cosmological ICs are fitted using  the Hubble $H$, deceleration $q$, jerk $j$, and snap $s$ parameters  evaluated at the present time $t=t_0$. We can adjust the first derivatives of the scale factor as  $a(0)=a_0=1\;, \dot{a}(0)=H_0a_0, \ddot{a}(0)=-H_0^2a_0q_0\;, \dddot{a}(0)=H_0^3 j_0 a_0^{-1}$ where $q_0$ is the deceleration parameter at the initial time (present time), $j_0$ is the jerk parameter at the instant $t=t_0$, etc \cite{visser04}.
Fortunately, these  data have been measured with high precisions.

\begin{figure}
\centering
 \includegraphics[scale=0.3] {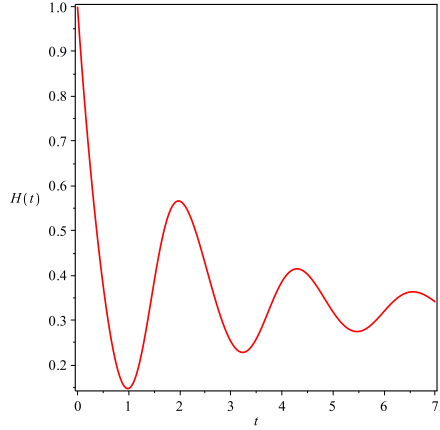}
  \caption{Numerical solution for  $H(t)$. The model of $f(R)$ is the one proposed by Starobinsky, with $\alpha=0.02$. The cosmological data are fitted with observational data for extended cosmological parameters.}
  \label{H.eps}
\end{figure}

A series solution for $a(t)$ in Eqn \eqref{efe1-1} has been developed using these cosmographic parameters which  are all evaluated at $t=t_0$: 
\begin{eqnarray}
&&a(t)=1+H_{{0}} \left( t-t_{{0}} \right) -1/2\,{H_{{0}}}^{2}q_{{0}} \left( t
-t_{{0}} \right) ^{2}\nn
&&-{\frac {1}{216}}\,{\frac { \left( -3\,{H_{{0}}}^
{2}+54\,{H_{{0}}}^{4}\alpha+\mu_{{m}}+12\,\alpha\,\mu_{{m}}{H_{{0}}}^{
2}-12\,\alpha\,\mu^0_{{m}}{H_{{0}}}^{2}q_{{0}}+18\,\alpha\,{H_{{0}}}^{4}
{q_{{0}}}^{2}+36\,{H_{{0}}}^{4}\alpha\,q_{{0}} \right)  }{\alpha\,H_{{0}}}}\left( t-t_{{0
}} \right) ^{3}\nn
&&+{\frac {1}{2592}}\,{\frac { \left( t-t_{{0}} \right) ^{4}}{\alpha\,{H_
{{0}}}^{2}}}\times\Big(9\,{H_{{0}}}^{4}+162\,{H_{{0}}}^{6}\alpha-12\,\mu^0_{{m}}{H_{{0}}}^{2}+
18\,{H_{{0}}}^{4}\alpha\,\mu^0_{{m}}+108\,{H_{{0}}}^{4}\alpha\,\mu^0_{{m}}
q_{{0}}-54\,{H_{{0}}}^{6}\alpha\,{q_{{0}}}^{2}\nn
&&+324\,{H_{{0}}}^{6}
\alpha\,q_{{0}}-108\,\alpha\,{H_{{0}}}^{6}{q_{{0}}}^{3}-6\,\mu^0_{{m}}{H
_{{0}}}^{2}q_{{0}}+9\,\mu^0_{{m}}\ga_{{m}}{H_{{0}}}^{2}+{\mu_{{m}}}^{2}+90
\,\alpha\,\mu^0_{{m}}{H_{{0}}}^{4}{q_{{0}}}^{2}-12\,\alpha\,{\mu^0_{{m}}}^
{2}{H_{{0}}}^{2}q_{{0}}\nn
&&+12\,\alpha\,{\mu_{{m}}}^{2}{H_{{0}}}^{2}+108\,
\mu^0_{{m}}\ga_{{m}}{H_{{0}}}^{4}\alpha-108\,\mu^0_{{m}}\ga_{{m}}{H_{{0}}}^{4}
\alpha\,q_{{0}}
\Big)+O\left[(t-t_0)^5\right]\;.
\end{eqnarray}
The above solution can be used to check observational constraints. As an alternative, we can also solve  Eqn  \eqref{efe1-1} numerically. 
A numerical solution for the Hubble parameter is developed in Fig. \ref{H.eps} where we put $a_0=H_0=1\;, q_0=-0.7$. 
We see in Fig. \ref{H.eps} that $H$  is an oscillatory function, it reaches maxima and minima several times. It defines an oscillatory solution  but it is not in the form of Type IV future singularity \cite{nojiri05,brand12,cai14, bamba08,barr04}.
But it can be identified in the late-time as the $\Lambda$CDM era.

\begin{figure}[htb!]
\centering
 \includegraphics[scale=0.4] {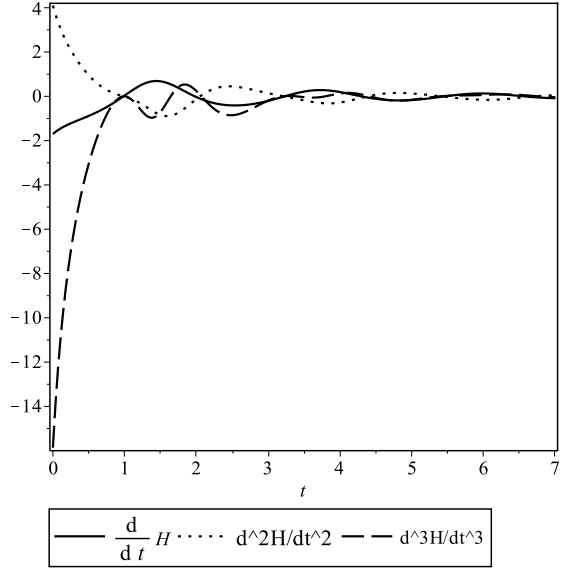}
  \caption{Numerical solution for  $\dot{H}\;,\ddot{H}\;,\dddot{H}$.}
  \label{H123.eps}
\end{figure}

We can classify the future singularities as follows:
\begin{itemize}
\item Type I: (``Big Rip"):   $t\to t_s$, $a\to \infty\;,\mu\to \infty$ and $|p| \to \infty$. 

\item  Type II: (``sudden"):  
$t\to t_s$,$a\to a_s\;,\mu\to \mu_s$ and $|p| \to \infty$. 

\item  Type III : 
$t\to t_s$, $a\to a_s\;,\mu\to \infty$ and $|p| \to \infty$

\item  Type IV : 
$t\to t_s$, $a\to a_s\;,\mu\to 0$ and $|p| \to 0$
 and
higher derivatives of H diverge.
Here $t_s\;, a_s$ and $\mu_s$ are constants with $a_s\neq0$.

\end{itemize}
For our case, the factor given in Fig. \ref{H.eps}, the Hubble parameter and first, second and third derivatives of $H$ are plotted in Fig. \ref{H123.eps}. No higher derivatives of H diverges.

\begin{figure}[htb!]
\centering
 \includegraphics[scale=0.4] {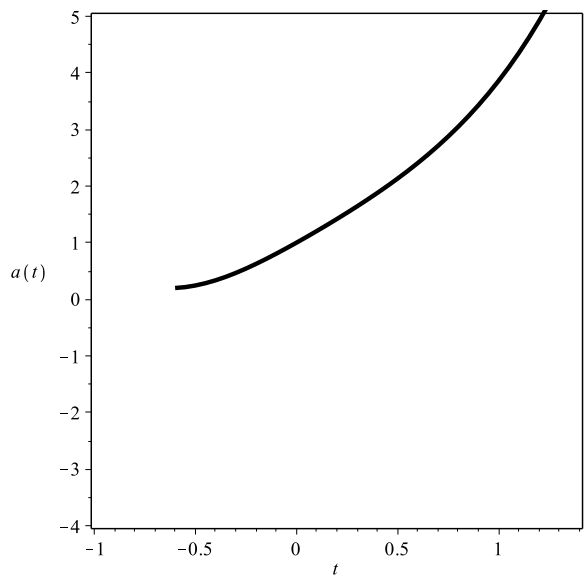}
  \caption{ Phase portrait for Starobinsky's model.}
  \label{phase1.eps}
\end{figure}

A phase portrait for Starobinsky model is plotted in Fig. \ref{phase1.eps}. Here we solved the ODE with parameters $\Omega^{0}_{m}\equiv \frac{\mu^{0}_{m}}{3H^{2}_{0}}=0.3\;, \ga_{m}=1$.
The phase portrait shows that the scale factor $a(t)$ is a monotonic increasing function of time. It is always increasing, and never decreasing.

For curiosity we are interested to know if the system has attractors or not. The late-time or asymptotic attractors are a class of solutions which have a generic form independent of the initial conditions. We examine our model for such types of solutions and solve the equations of motion for some initial conditions. The model is well established as an attractor in the following Fig. \ref{attractor.eps}.
\begin{figure}[htb!]
\centering
 \includegraphics[scale=0.4] {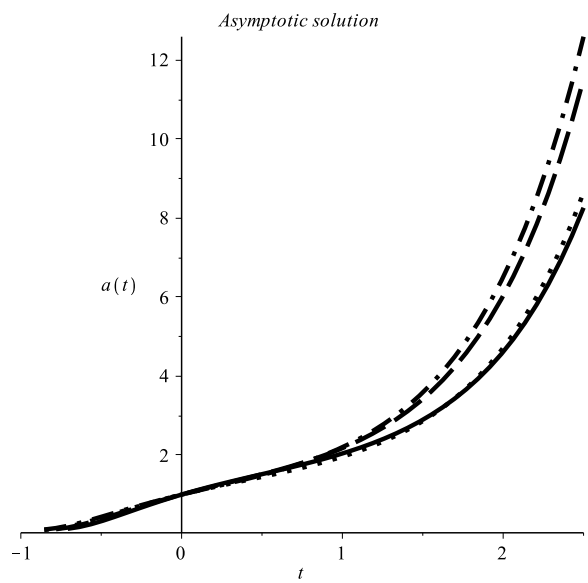}
  \caption{Attractors for Starobinsky's model. }
  \label{attractor.eps}
\end{figure}

\np
\section{Discussions and Conclusion}\label{conc}
In this work we looked at classes of shear-free anisotropic cosmological spacetimes in $f(R)$ gravity. Focusing on orthogonal models with irrotational and non-accelerated fluid flows without heat fluxes, we showed that the anisotropic stresses are related to the electric part of the Weyl tensor in such a way that they balance each other. This is considered necessary and sufficient condition for the shear to be vanishing forever if vanishing initially. This turned out to be a generalization of a previous result \cite{mimo93} for models whose underlying theory is GR. We also showed that within the class of orthogonal $f(R)$ models, small perturbations of shear are damped, i.e,. that these models are stable if expanding, and that the electric part of the Weyl tensor and the anisotropic stress tensor decay with the expansion as well as the heat flux of the curvature fluid.

 As an application, we considered a subclass of locally rotationally symmetric spacetimes with barotropic equations of state and studied the evolutionary dynamics of the Universe. In particular, we showed that the late-time behaviour of the dS universe in $f(R)$ gravity should satisfy Eqn \eqref{dS}. For the Starobinsky model of $f(R)$, we provided a power-series solution for $a(t)$ and we studied the behavior of the expansion parameter $H(t)$  by numerically integrating the Friedmann equation \eqref{efe1-1}, where the intial conditions for $H_{0}\;, q_{0}$ and $j_{0}$ are taken from observational data. The result is the oscillatory solution presented in Fig. \ref{H.eps} and describes the late-time universe in the $\Lambda$CDM era. The first three derivatives of H have also been calculated as shown in Fig. \ref{H123.eps}; none of these derivatives diverges. A phase-portrait anaysis for this model with $\Omega^{0}_{m}=0.3\;, \ga_{m}=1$, given in Fig. \ref{phase1.eps}, shows that the scale factor is a monotonically increasing function of time. Finally, we examined our model for late-time or asymptotic attractors, with well established solutions depicted in Fig. \ref{attractor.eps}.

%
%


\bibliography{bibliography}

\begin{thebibliography}{10}
\expandafter\ifx\csname url\endcsname\relax
  \def\url#1{\texttt{#1}}\fi
\expandafter\ifx\csname urlprefix\endcsname\relax\def\urlprefix{URL }\fi
\providecommand{\bibinfo}[2]{#2}
\providecommand{\eprint}[2][]{\url{#2}}

\bibitem{capozziello11extended}
\bibinfo{author}{Capozziello, S.} \& \bibinfo{author}{De~Laurentis, M.}
\newblock \bibinfo{title}{Extended theories of gravity}.
\newblock \emph{\bibinfo{journal}{Physics Reports}}
  \textbf{\bibinfo{volume}{509}}, \bibinfo{pages}{167--321}
  (\bibinfo{year}{2011}).

\bibitem{modesto12}
\bibinfo{author}{Modesto, L.}
\newblock \bibinfo{title}{Super-renormalizable quantum gravity}.
\newblock \emph{\bibinfo{journal}{Phys. Rev. D}} \textbf{\bibinfo{volume}{86}},
  \bibinfo{pages}{044005} (\bibinfo{year}{2012}).

\bibitem{modesto14}
\bibinfo{author}{Modesto, L.} \& \bibinfo{author}{Rachwal, L.}
\newblock \bibinfo{title}{Super-renormalizable and finite gravitational
  theories}.
\newblock \emph{\bibinfo{journal}{Nuclear Physics B}}
  \textbf{\bibinfo{volume}{889}}, \bibinfo{pages}{228--248}
  (\bibinfo{year}{2014}).

\bibitem{biswas12}
\bibinfo{author}{Biswas, T.}, \bibinfo{author}{Gerwick, E.},
  \bibinfo{author}{Koivisto, T.} \& \bibinfo{author}{Mazumdar, A.}
\newblock \bibinfo{title}{{Towards singularity and ghost free theories of
  gravity}}.
\newblock \emph{\bibinfo{journal}{Phys.Rev.Lett.}}
  \textbf{\bibinfo{volume}{108}}, \bibinfo{pages}{031101}
  (\bibinfo{year}{2012}).

\bibitem{clifton12}
\bibinfo{author}{Clifton, T.}, \bibinfo{author}{Ferreira, P.~G.},
  \bibinfo{author}{Padilla, A.} \& \bibinfo{author}{Skordis, C.}
\newblock \bibinfo{title}{Modified gravity and cosmology}.
\newblock \emph{\bibinfo{journal}{Physics Reports}}
  \textbf{\bibinfo{volume}{513}}, \bibinfo{pages}{1--189}
  (\bibinfo{year}{2012}).

\bibitem{de2010f}
\bibinfo{author}{De~Felice, A.} \& \bibinfo{author}{Tsujikawa, S.}
\newblock \bibinfo{title}{${f(R)}$ theories}.
\newblock \emph{\bibinfo{journal}{Living Rev. Rel}}
  \textbf{\bibinfo{volume}{13}}, \bibinfo{pages}{1002--4928}
  (\bibinfo{year}{2010}).

\bibitem{noj07}
\bibinfo{author}{Nojiri, S.} \& \bibinfo{author}{Odintsov, S.~D.}
\newblock \bibinfo{title}{Introduction to modified gravity and gravitational
  alternative for dark energy}.
\newblock \emph{\bibinfo{journal}{International Journal of Geometric Methods in
  Modern Physics}} \textbf{\bibinfo{volume}{4}}, \bibinfo{pages}{115--145}
  (\bibinfo{year}{2007}).

\bibitem{noj11}
\bibinfo{author}{Nojiri, S.} \& \bibinfo{author}{Odintsov, S.~D.}
\newblock \bibinfo{title}{Unified cosmic history in modified gravity: from f
  (r) theory to lorentz non-invariant models}.
\newblock \emph{\bibinfo{journal}{Physics Reports}}
  \textbf{\bibinfo{volume}{505}}, \bibinfo{pages}{59--144}
  (\bibinfo{year}{2011}).

\bibitem{buchdal70}
\bibinfo{author}{{Buchdahl}, H.~A.}
\newblock \bibinfo{title}{{Non-linear Lagrangians and cosmological theory}}.
\newblock \emph{\bibinfo{journal}{Monthly Notices of the Royal Astronomical
  Society}} \textbf{\bibinfo{volume}{150}}, \bibinfo{pages}{1}
  (\bibinfo{year}{1970}).

\bibitem{staro80}
\bibinfo{author}{{Starobinsky}, A.~A.}
\newblock \bibinfo{title}{{A new type of isotropic cosmological models without
  singularity}}.
\newblock \emph{\bibinfo{journal}{Physics Letters B}}
  \textbf{\bibinfo{volume}{91}}, \bibinfo{pages}{99--102}
  (\bibinfo{year}{1980}).

\bibitem{carroll04}
\bibinfo{author}{Carroll, S.}, \bibinfo{author}{Duvvuri, V.},
  \bibinfo{author}{Turner, M.} \& \bibinfo{author}{Trodden, M.}
\newblock \bibinfo{title}{Is cosmic speed-up due to new gravitational physics?}
\newblock \emph{\bibinfo{journal}{Physical Review D}}
  \textbf{\bibinfo{volume}{70}}, \bibinfo{pages}{043528}.

\bibitem{noj03}
\bibinfo{author}{Nojiri, S.} \& \bibinfo{author}{Odintsov, S.~D.}
\newblock \bibinfo{title}{Modified gravity with negative and positive powers of
  curvature: Unification of inflation and cosmic acceleration}.
\newblock \emph{\bibinfo{journal}{Physical Review D}}
  \textbf{\bibinfo{volume}{68}}, \bibinfo{pages}{123512}
  (\bibinfo{year}{2003}).

\bibitem{sotiriou07}
\bibinfo{author}{Sotiriou, T.~P.} \& \bibinfo{author}{Liberati, S.}
\newblock \bibinfo{title}{Metric-affine ${f(R)}$ theories of gravity}.
\newblock \emph{\bibinfo{journal}{Annals of Physics}}
  \textbf{\bibinfo{volume}{322}}, \bibinfo{pages}{935--966}
  (\bibinfo{year}{2007}).

\bibitem{nojiri06}
\bibinfo{author}{Nojiri, S.} \& \bibinfo{author}{Odintsov, S.~D.}
\newblock \bibinfo{title}{Modified ${f(R)}$ gravity consistent with realistic
  cosmology: From a matter dominated epoch to a dark energy universe}.
\newblock \emph{\bibinfo{journal}{Physical Review D}}
  \textbf{\bibinfo{volume}{74}}, \bibinfo{pages}{086005}
  (\bibinfo{year}{2006}).

\bibitem{staro07}
\bibinfo{author}{Starobinsky, A.~A.}
\newblock \bibinfo{title}{Disappearing cosmological constant in ${f(R)}$
  gravity}.
\newblock \emph{\bibinfo{journal}{JETP Letters}} \textbf{\bibinfo{volume}{86}},
  \bibinfo{pages}{157--163} (\bibinfo{year}{2007}).

\bibitem{mimo93}
\bibinfo{author}{Mimoso, J.~P.} \& \bibinfo{author}{Crawford, P.}
\newblock \bibinfo{title}{Shear-free anisotropic cosmological models}.
\newblock \emph{\bibinfo{journal}{Classical and Quantum Gravity}}
  \textbf{\bibinfo{volume}{10}}, \bibinfo{pages}{315} (\bibinfo{year}{1993}).

\bibitem{ellis67}
\bibinfo{author}{Ellis, G.}
\newblock \bibinfo{title}{Dynamics of pressure-free matter in general
  relativity}.
\newblock \emph{\bibinfo{journal}{Journal of Mathematical Physics}}
  \textbf{\bibinfo{volume}{8}}, \bibinfo{pages}{1171} (\bibinfo{year}{1967}).

\bibitem{godel52}
\bibinfo{author}{G{\"o}del, K.}
\newblock \bibinfo{title}{Rotating universes in general relativity theory}.
\newblock In \emph{\bibinfo{booktitle}{Proceedings of the International
  Congress of Mathematicians Edited by LM Graves et al., Cambridge, Mass. 1952,
  vol. 1, p. 175.}}, vol.~\bibinfo{volume}{1}, \bibinfo{pages}{175}
  (\bibinfo{year}{1952}).

\bibitem{goldberg62}
\bibinfo{author}{Goldberg, J.} \& \bibinfo{author}{Sachs, R.}
\newblock \bibinfo{title}{A theorem on petrov type(field equations for proving
  theorem identifying geometrical properties of null congruence with existence
  of algebraically special riemann tensor)}.
\newblock \emph{\bibinfo{journal}{1966.}} \bibinfo{pages}{13--23}
  (\bibinfo{year}{1962}).

\bibitem{robinson63}
\bibinfo{author}{Robinson, I.} \& \bibinfo{author}{Schild, A.}
\newblock \bibinfo{title}{Generalization of a theorem by goldberg and sachs}.
\newblock \emph{\bibinfo{journal}{Journal of Mathematical Physics}}
  \textbf{\bibinfo{volume}{4}}, \bibinfo{pages}{484} (\bibinfo{year}{1963}).

\bibitem{ellis69}
\bibinfo{author}{Ellis, G.} \& \bibinfo{author}{MacCallum, M.~A.}
\newblock \bibinfo{title}{A class of homogeneous cosmological models}.
\newblock \emph{\bibinfo{journal}{Communications in Mathematical Physics}}
  \textbf{\bibinfo{volume}{12}}, \bibinfo{pages}{108--141}
  (\bibinfo{year}{1969}).

\bibitem{macc71}
\bibinfo{author}{MacCallum, M.}
\newblock \bibinfo{title}{A class of homogeneous cosmological models iii:
  asymptotic behaviour}.
\newblock \emph{\bibinfo{journal}{Communications in Mathematical Physics}}
  \textbf{\bibinfo{volume}{20}}, \bibinfo{pages}{57--84}
  (\bibinfo{year}{1971}).

\bibitem{collins86}
\bibinfo{author}{Collins, C.}
\newblock \bibinfo{title}{Shear-free fluids in general relativity}.
\newblock \emph{\bibinfo{journal}{Canadian journal of physics}}
  \textbf{\bibinfo{volume}{64}}, \bibinfo{pages}{191--199}
  (\bibinfo{year}{1986}).

\bibitem{barrow77}
\bibinfo{author}{Barrow, J.~D.} \& \bibinfo{author}{Matzner, R.~A.}
\newblock \bibinfo{title}{The homogeneity and isotropy of the universe}.
\newblock \emph{\bibinfo{journal}{Monthly Notices of the Royal Astronomical
  Society}} \textbf{\bibinfo{volume}{181}}, \bibinfo{pages}{719--727}
  (\bibinfo{year}{1977}).

\bibitem{macc70}
\bibinfo{author}{MacCallum, M.}, \bibinfo{author}{Stewart, J.} \&
  \bibinfo{author}{Schmidt, B.}
\newblock \bibinfo{title}{Anisotropic stresses in homogeneous cosmologies}.
\newblock \emph{\bibinfo{journal}{Communications in Mathematical Physics}}
  \textbf{\bibinfo{volume}{17}}, \bibinfo{pages}{343--347}
  (\bibinfo{year}{1970}).

\bibitem{koi11}
\bibinfo{author}{Koivisto, T.~S.}, \bibinfo{author}{Mota, D.~F.},
  \bibinfo{author}{Quartin, M.} \& \bibinfo{author}{Zlosnik, T.~G.}
\newblock \bibinfo{title}{Possibility of anisotropic curvature in cosmology}.
\newblock \emph{\bibinfo{journal}{Physical Review D}}
  \textbf{\bibinfo{volume}{83}}, \bibinfo{pages}{023509}
  (\bibinfo{year}{2011}).

\bibitem{barr83}
\bibinfo{author}{Barrow, J.~D.} \& \bibinfo{author}{Ottewill, A.~C.}
\newblock \bibinfo{title}{The stability of general relativistic cosmological
  theory}.
\newblock \emph{\bibinfo{journal}{Journal of Physics A: Mathematical and
  General}} \textbf{\bibinfo{volume}{16}}, \bibinfo{pages}{2757}
  (\bibinfo{year}{1983}).

\bibitem{barr06}
\bibinfo{author}{Barrow, J.~D.} \& \bibinfo{author}{Clifton, T.}
\newblock \bibinfo{title}{Exact cosmological solutions of scale-invariant
  gravity theories}.
\newblock \emph{\bibinfo{journal}{Classical and Quantum Gravity}}
  \textbf{\bibinfo{volume}{23}}, \bibinfo{pages}{L1} (\bibinfo{year}{2006}).

\bibitem{clift06}
\bibinfo{author}{Clifton, T.} \& \bibinfo{author}{Barrow, J.~D.}
\newblock \bibinfo{title}{Further exact cosmological solutions to higher-order
  gravity theories}.
\newblock \emph{\bibinfo{journal}{Classical and Quantum Gravity}}
  \textbf{\bibinfo{volume}{23}}, \bibinfo{pages}{2951} (\bibinfo{year}{2006}).

\bibitem{midd10}
\bibinfo{author}{Middleton, J.}
\newblock \bibinfo{title}{On the existence of anisotropic cosmological models
  in higher order theories of gravity}.
\newblock \emph{\bibinfo{journal}{Classical and Quantum Gravity}}
  \textbf{\bibinfo{volume}{27}}, \bibinfo{pages}{225013}
  (\bibinfo{year}{2010}).

\bibitem{abebe11}
\bibinfo{author}{Abebe, A.}, \bibinfo{author}{Goswami, R.} \&
  \bibinfo{author}{Dunsby, P.}
\newblock \bibinfo{title}{{Shear-free perturbations of ${f(R)}$ gravity}}.
\newblock \emph{\bibinfo{journal}{Physical Review D}} \bibinfo{pages}{1--7}.

\bibitem{abebe2013}
\bibinfo{author}{Abebe, A.}
\newblock \emph{\bibinfo{title}{Beyond concordance cosmology}}.
\newblock Ph.D. thesis, \bibinfo{school}{UCT (University of Cape Town)}
  (\bibinfo{year}{2013}).

\bibitem{amare14a}
\bibinfo{author}{Abebe, A.}
\newblock \bibinfo{title}{Anti-newtonian cosmologies in ${f(R)}$ gravity}.
\newblock \emph{\bibinfo{journal}{Classical and Quantum Gravity}}
  \textbf{\bibinfo{volume}{31}}, \bibinfo{pages}{115011}
  (\bibinfo{year}{2014}).

\bibitem{maye14}
\bibinfo{author}{Abebe, A.} \& \bibinfo{author}{Elmardi, M.}
\newblock \bibinfo{title}{Irrotational-fluid cosmologies in fourth-order
  gravity}.
\newblock \emph{\bibinfo{journal}{arXiv preprint arXiv:1411.6394}}
  (\bibinfo{year}{2014}).

\bibitem{Ellis98}
\bibinfo{author}{Ellis, G.} \& \bibinfo{author}{van Elst, H.}
\newblock \bibinfo{title}{Cosmological models}.
\newblock In \emph{\bibinfo{booktitle}{Theoretical and Observational
  Cosmology}}, \bibinfo{pages}{1--116} (\bibinfo{publisher}{Dordrecht: Kluver},
  \bibinfo{year}{1999}).

\bibitem{maartens97}
\bibinfo{author}{Maartens, R.} \& \bibinfo{author}{Triginer, J.}
\newblock \bibinfo{title}{Density perturbations with relativistic
  thermodynamics}.
\newblock \emph{\bibinfo{journal}{Physical Review D}}
  \textbf{\bibinfo{volume}{56}}, \bibinfo{pages}{4640} (\bibinfo{year}{1997}).

\bibitem{betschart}
\bibinfo{author}{Betschart, G.}
\newblock \emph{\bibinfo{title}{{General relativistic electrodynamics with
  applicantions in cosmology and astrophysics}}}.
\newblock Ph.D. thesis, \bibinfo{school}{University of Cape Town}
  (\bibinfo{year}{2005}).

\bibitem{carloni08}
\bibinfo{author}{Carloni, S.}, \bibinfo{author}{Dunsby, P.} \&
  \bibinfo{author}{Troisi, A.}
\newblock \bibinfo{title}{Evolution of density perturbations in {${f(R)}$}
  gravity}.
\newblock \emph{\bibinfo{journal}{Physical Review D}}
  \textbf{\bibinfo{volume}{77}}, \bibinfo{pages}{024024}
  (\bibinfo{year}{2008}).

\bibitem{ellis89}
\bibinfo{author}{Ellis, G.} \& \bibinfo{author}{Bruni, M.}
\newblock \bibinfo{title}{Covariant and gauge-invariant approach to
  cosmological density fluctuations}.
\newblock \emph{\bibinfo{journal}{Physical Review D}}
  \textbf{\bibinfo{volume}{40}}, \bibinfo{pages}{1804} (\bibinfo{year}{1989}).

\bibitem{ellis12}
\bibinfo{author}{Ellis, G.}, \bibinfo{author}{Maartens, R.} \&
  \bibinfo{author}{MacCallum, M.~A.}
\newblock \emph{\bibinfo{title}{Relativistic cosmology}}
  (\bibinfo{publisher}{Cambridge University Press}, \bibinfo{year}{2012}).

\bibitem{maartens98}
\bibinfo{author}{Maartens, R.}
\newblock \bibinfo{title}{Covariant velocity and density perturbations in
  quasi-newtonian cosmologies}.
\newblock \emph{\bibinfo{journal}{Physical Review D}}
  \textbf{\bibinfo{volume}{58}}, \bibinfo{pages}{124006}
  (\bibinfo{year}{1998}).

\bibitem{wern76}
\bibinfo{author}{Israel, W.}
\newblock \bibinfo{title}{Nonstationary irreversible thermodynamics: A causal
  relativistic theory}.
\newblock \emph{\bibinfo{journal}{Annals of Physics}}
  \textbf{\bibinfo{volume}{100}} (\bibinfo{year}{1976}).
\newblock
  \urlprefix\url{http://gen.lib.rus.ec/scimag/index.php?s=10.1016/0003-4916(76)90064-6}.

\bibitem{novella96}
\bibinfo{author}{Novella, M.}
\newblock \emph{\bibinfo{title}{Cosmology and Gravitation Two}},
  vol.~\bibinfo{volume}{2} (\bibinfo{publisher}{Atlantica S{\'e}guier
  Fronti{\`e}res}, \bibinfo{year}{1996}).

\bibitem{maartens96}
\bibinfo{author}{Maartens, R.}
\newblock \bibinfo{title}{Causal thermodynamics in relativity}.
\newblock \emph{\bibinfo{journal}{arXiv preprint astro-ph/9609119}}
  (\bibinfo{year}{1996}).

\bibitem{rezz13}
\bibinfo{author}{Rezzolla, L.} \& \bibinfo{author}{Zanotti, O.}
\newblock \emph{\bibinfo{title}{Relativistic hydrodynamics}}
  (\bibinfo{publisher}{Oxford University Press}, \bibinfo{year}{2013}).

\bibitem{col95}
\bibinfo{author}{Coley, A.~A.} \& \bibinfo{author}{Van~den Hoogen, R.}
\newblock \bibinfo{title}{Qualitative analysis of viscous fluid cosmological
  models satisfying the israel-stewart theory of irreversible thermodynamics}.
\newblock \emph{\bibinfo{journal}{Classical and Quantum Gravity}}
  \textbf{\bibinfo{volume}{12}}, \bibinfo{pages}{1977} (\bibinfo{year}{1995}).

\bibitem{visser04}
\bibinfo{author}{Visser, M.}
\newblock \bibinfo{title}{Jerk, snap and the cosmological equation of state}.
\newblock \emph{\bibinfo{journal}{Classical and Quantum Gravity}}
  \textbf{\bibinfo{volume}{21}}, \bibinfo{pages}{2603} (\bibinfo{year}{2004}).

\bibitem{nojiri05}
\bibinfo{author}{Nojiri, S.}, \bibinfo{author}{Odintsov, S.} \&
  \bibinfo{author}{Tsujikawa, S.}
\newblock \bibinfo{title}{Properties of singularities in (phantom) dark energy
  universe, p hys. rev. d71(2005) 063004}.
\newblock \emph{\bibinfo{journal}{arXiv preprint hep-th/0501025}} .

\bibitem{brand12}
\bibinfo{author}{Brandenberger, R.~H.}
\newblock \bibinfo{title}{The matter bounce alternative to inflationary
  cosmology}.
\newblock \emph{\bibinfo{journal}{arXiv preprint arXiv:1206.4196}}
  (\bibinfo{year}{2012}).

\bibitem{cai14}
\bibinfo{author}{Cai, Y.-F.}
\newblock \bibinfo{title}{Exploring bouncing cosmologies with cosmological
  surveys}.
\newblock \emph{\bibinfo{journal}{Science China Physics, Mechanics \&
  Astronomy}} \textbf{\bibinfo{volume}{57}}, \bibinfo{pages}{1414--1430}
  (\bibinfo{year}{2014}).

\bibitem{bamba08}
\bibinfo{author}{Bamba, K.}, \bibinfo{author}{Nojiri, S.} \&
  \bibinfo{author}{Odintsov, S.~D.}
\newblock \bibinfo{title}{The future of the universe in modified gravitational
  theories: approaching a finite-time future singularity}.
\newblock \emph{\bibinfo{journal}{Journal of Cosmology and Astroparticle
  Physics}} \textbf{\bibinfo{volume}{2008}}, \bibinfo{pages}{045}
  (\bibinfo{year}{2008}).

\bibitem{barr04}
\bibinfo{author}{Barrow, J.~D.}
\newblock \bibinfo{title}{More general sudden singularities}.
\newblock \emph{\bibinfo{journal}{Classical and Quantum Gravity}}
  \textbf{\bibinfo{volume}{21}}, \bibinfo{pages}{5619} (\bibinfo{year}{2004}).

\end{thebibliography}
\bibliographystyle{naturemag}

\end{document}